# Design and experimental demonstration of photonic-crystal lasers with multijunction active layers

Shumpei Katsuno[1*†], Masahiro Yoshida[1†], Takuya Inoue[2†], Menaka De Zoysa[2], Ranko Hatsuda[1], Kenji Ishizaki[2], and Susumu Noda[1,2*]

[1] Department of Electronic Science and Engineering, Kyoto University,
[2] Photonics and Electronics Science and Engineering Center, Kyoto University,
Kyoto 615-8510, Japan

*E-mail: s.katsuno@qoe.kuee.kyoto-u.ac.jp, snoda@kuee.kyoto-u.ac.jp
[†]These authors contributed equally to this work.

**Abstract**

**We introduce multijunction active layers, featuring a stack of alternating active layers and tunnel junctions, to PCSELs to increase their slope efficiency, which is vital for various applications including laser processing and LiDAR. First, we design a multijunction PCSEL that avoids optical absorption in the heavily-doped tunnel junctions while allowing sufficient optical gain and resonance effects in the active and photonic crystal layers. Next, we fabricate a 3-mm-diameter two-junction PCSEL, achieving a slope efficiency of 1.58 W/A, which is over twice as high as that of conventional single-junction PCSELs, and a record-high peak output power of 1.8 kW for PCSELs.**

Photonic-crystal surface-emitting lasers (PCSELs) are attracting much attention as next-generation semiconductor lasers capable of high-power and high-beam-quality, namely high-brightness, operation, which is difficult to realize with conventional semiconductor lasers, owing to large-area coherent oscillation via resonance at singularity (e.g., Γ, M) points in two-dimensional (2D) photonic crystals[1–11]. Recently, based on the use of a double-lattice structure[12] to appropriately control optical couplings between light waves inside the photonic crystal (more precisely, Hermitian coupling, which is not accompanied by optical loss, and non-Hermitian coupling, which is accompanied by radiation losses), a design guideline has been established for realizing single-mode oscillation over large areas with circular diameters of 3 mm or more[13]. Furthermore, based on this guideline, single-mode operation with a high output power of 50 W under continuous-wave (CW) conditions has been experimentally achieved with a 3-mm-diameter PCSEL, resulting in a high brightness of 1 GWcm$^{-2}$sr$^{-1}$, rivalling that of large lasers ($CO_2$ lasers and solid-state lasers including fiber lasers)[14]. In addition, even higher output powers can be obtained under pulsed operation of such a large-area PCSEL; as a recent example, a peak output power of 1 kW has been achieved while maintaining a brightness of 1 GWcm$^{-2}$sr$^{-1}$ [15,16]. Besides high-brightness operation, PCSELs can also achieve high functionalities, such as electronic 2D beam scanning and the generation of various (e.g., flash and multi-dot) beam patterns[16-18]. Such functionalities are achieved by introducing photonic crystals in which the position and/or size of air holes in the photonic crystal are modulated. Due to the above features, PCSELs are expected to contribute a wide range of applications in fields such as LiDAR, laser processing, and aerospace (including free-space communications and space-based LiDAR).

For the various applications described above, it is important to increase the slope efficiency of PCSELs, i.e., the rate of change of optical output power with respect to injection current, in order to increase the output power at a given injection current or, alternatively, to decrease the injection current at a given output power. To this end, we have proposed the introduction of multijunction active layers[19,20]. While conventional PCSELs have a single active layer sandwiched between n- and p-type cladding layers, multijunction PCSELs have a stack of alternating active layers and tunnel-junction (i.e., highly doped p-n junction) layers, allowing a single set of electron-hole pairs injected into the device from the n- and p-side electrodes to lead to the emission of multiple photons by radiative recombination in multiple active layers. As a result, the slope efficiency is expected to increase in proportion to the number of stacked active layers.

In this paper, we design and experimentally demonstrate a PCSEL with multijunction active layers. First, we design a device structure of a multijunction PCSEL that enables sufficient optical confinement in the active and photonic crystal layers to obtain sufficient optical gain and resonance effects, while avoiding absorption losses in the heavily-doped tunnel-junction layer. Next, we simulate the lasing characteristics of the designed PCSEL with two-junction active layers and show that the slope efficiency can be increased while maintaining high-beam-quality operation based on the band-edge resonance effect of the photonic crystal. Furthermore, based on this design, we fabricate a 3-mm-diameter two-junction PCSEL and experimentally demonstrate a high slope efficiency of 1.58 W/A, which is over twice as high as that of a conventional single-junction PCSEL, and a record-high peak optical power of 1.8 kW for PCSELs under pulsed condition.

Figure 1(a) shows a schematic of the layer structure of a PCSEL with multijunction active layers. As in a conventional PCSEL, the active layers and the photonic crystal layer are placed between n- and p-type cladding layers. However, unlike in a conventional PCSEL, the multiple active layers (two layers in this figure) are stacked via heavily-doped $p^{++}$-$n^{++}$ tunnel-junction layer(s), where the order of the p- and n-type layers is reversed[19,20]. Figure 1(b) shows a schematic of an energy band diagram near the active layers of a two-junction PCSEL under forward bias, where tunnel current flows owing to the transfer of electrons from the p-side valence band to the n-side conduction band. As a result, when an electron-hole pair is injected through the electrode, multiple photons (two photons in this case) are emitted. In this way, the introduction of multijunction active layers is expected to increase the slope efficiency.

It should be noted that since the tunnel junctions are heavily doped, a large optical confinement factor in the tunnel layers ($\Gamma_{tunnel}$) would increase internal material loss due to free-carrier absorption. To avoid this, the vertical electrical-field distribution of the waveguide mode, which is vertically confined to the higher-refractive-index active and photonic crystal layers between the lower-refractive-index cladding layers by total internal reflection, must be properly controlled. Generally, a single, fundamental waveguide mode, which has an antinode near the active layer and the photonic-crystal layer, is used for oscillation. However, for PCSELs with a multijunction active layer, this fundamental mode has a large $\Gamma_{tunnel}$, resulting in an increase of internal material loss in the tunnel junctions. Therefore, we consider utilization of a higher-order waveguide mode. If the higher-order waveguide mode has nodes in the tunnel junctions and antinodes in the active and photonic crystal layers, then it would be possible to avoid the internal material loss caused by the tunnel junctions while obtaining optical gain and resonance effects sufficient for oscillation.

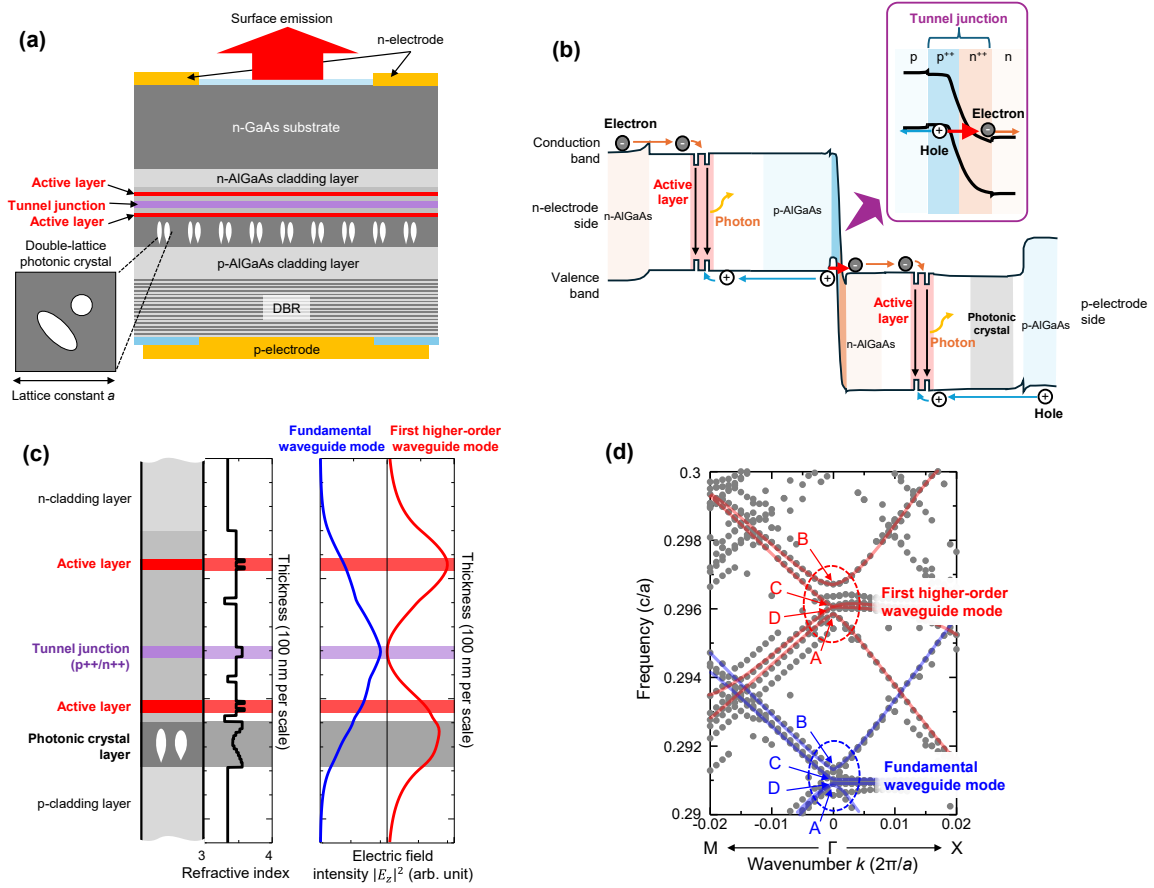

**Fig. 1.** Design of PCSELs with multijunction active layers. (a) Schematic of the device structure of a PCSEL with multijunction active layers (two active layers in this case). (b) Schematic of the energy band diagram near the active layers of a PCSEL with two-junction active layers. (c) Cross-sectional refractive index distribution of the designed layer structure with two-junction active layers (left), as well as calculated electric-field intensities of the fundamental and first higher-order waveguide modes (right). (d) Calculated photonic band diagram of a PCSEL with two-junction active layers.

**Table I.** Calculated optical confinement factors in the designed layer structure of a PCSEL with two-junction active layers.

| Waveguide mode | Tunnel junction ($\Gamma_{tunnel}$) | Active layers ($\Gamma_{act}$) | Photonic crystal layer ($\Gamma_{pc}$) |
|---|---|---|---|
| First higher-order waveguide mode | 0.01% | 4.85% | 16.81% |
| Fundamental waveguide mode | 6.41% | 3.87% | 7.88% |

Based on the above guideline, we designed a layer structure for a two-junction PCSEL (i.e., a PCSEL with two-junction active layers). For the photonic crystal layer, we considered a double-lattice photonic crystal with three-dimensional air hole shapes designed for operation at the Γ-point band edge. Figure 1(c) shows the cross-sectional refractive index distribution of the designed layer structure, as well as the electric field intensity $|E_z|^2$ of the fundamental and first higher-order waveguide modes calculated by the transfer matrix method. As shown in this figure, the fundamental waveguide mode has an antinode in the tunnel junction, while the first higher-order waveguide mode has antinodes in the active layers and photonic crystal layer and a node in the tunnel junction. Table I shows the optical confinement factors of the tunnel junction ($\Gamma_{tunnel}$), active layers ($\Gamma_{act}$), and photonic crystal layer ($\Gamma_{PC}$) for the fundamental and first higher-order waveguide modes. For the fundamental mode, $\Gamma_{tunnel}$ is large at 6.41%, resulting in large internal material loss. On the other hand, for the first higher-order waveguide mode, $\Gamma_{tunnel}$ is much smaller at 0.01%, while $\Gamma_{act}$ and $\Gamma_{PC}$ are 4.85% and 16.81%, respectively. Accordingly, PCSELs with this layer structure are expected to operate in the single, first higher-order waveguide mode with low internal material loss and sufficient optical gain and resonance effects.

Figure 1(d) shows the photonic band diagram near the Γ point for the above structure calculated by the rigorous coupled-wave analysis (RCWA) method. There exist two sets of four band edges (A, B, C, and D) at the Γ point. The sets at lower and higher frequencies correspond to the fundamental and first higher-order waveguide modes, respectively, reflecting the difference of the effective refractive index of each mode. High-beam-quality and high-slope-efficiency operation at the band edge of the first higher-order waveguide mode can be realized by adjusting the gain peak of the active layers to the frequency (wavelength) of this mode, in addition to obtaining sufficient optical gain and resonance effects as described above.

Next, we simulated the lasing characteristics of the two-junction PCSEL under pulsed operation based on the above design by using time-dependent three-dimensional coupled-wave theory[21]. Here, we assumed a resonator diameter of 3 mm and a double-lattice photonic-crystal structure with Hermitian and non-Hermitian coupling coefficients appropriate for 3-mm-diameter PCSELs as reported in Ref. 14 (i.e., $R$=24 cm$^1$, $I$=14 cm$^{-1}$, and $\mu$=44 cm$^{-1}$). Figure 2(a) shows the calculated *I-L* characteristics of the two-junction PCSEL (red) together with those of a conventional single-junction PCSEL for reference (blue). The slope efficiency of the two-junction PCSEL is almost twice as large as that of the single-junction PCSEL. Figures 2(b) and 2(c) show the calculated in-plane photon density distribution and far-field pattern of the two-junction PCSEL under an injection current of 110A. Owing to oscillation in a fundamental lateral mode, the two-junction PCSEL possesses a single-lobed in-plane photon density distribution and a narrow divergence of <0.05° without being affected by the vertically two-lobed profile of the first higher-order waveguide mode.

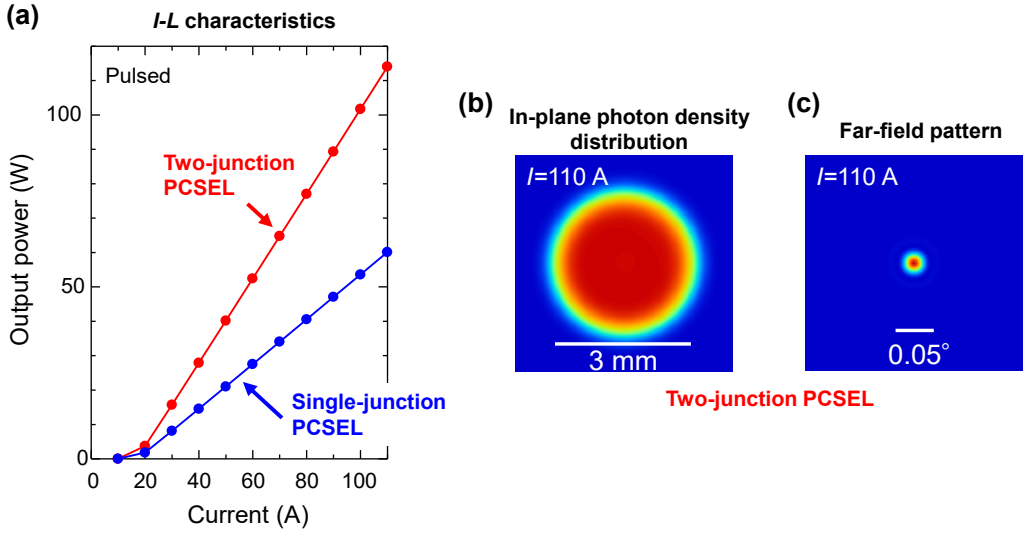

**Fig. 2.** Calculated lasing characteristics of a two-junction PCSEL under pulsed operation. (a) *I-L* characteristics of a two-junction PCSEL (red). *I-L* characteristics of a single-junction PCSEL are also shown for reference (blue). In these calculations, absorption of radiated light in the n-GaAs substrate and partial blocking of this light by the mesh electrode on the emission side are also considered. (b) In-plane photon density distribution and (c) far-field pattern of the two-junction PCSEL under an injection current of 110 A.

Based on the above design and simulation, we fabricated a two-junction PCSEL with a circular resonator area of 3 mm in diameter. First, we grew the n-cladding layer, active layers, tunnel-junction layer, and photonic-crystal layer by metal-organic vapor phase epitaxy (MOVPE), as shown in Fig.1 (a). Then, photonic-crystal patterns were fabricated by electron beam (EB) lithography and dry etching. After that, we grew a p-cladding layer, a distributed Bragg reflector (DBR) layer, and a p-contact layer by MOVPE regrowth, then we deposited electrodes onto the n and p sides. Figure 3(a) shows a microscope image of the fabricated two-junction PCSEL. A mesh electrode was deposited onto the n-GaAs substrate (emission side) for uniform current injection across the entire 3-mm-diameter area. Figure 3(b) shows a cross-sectional SEM image of the two-junction PCSEL together with a magnified image alongside a plot of the designed refractive index distribution near the active layers, indicating that the layer structure was fabricated as designed.

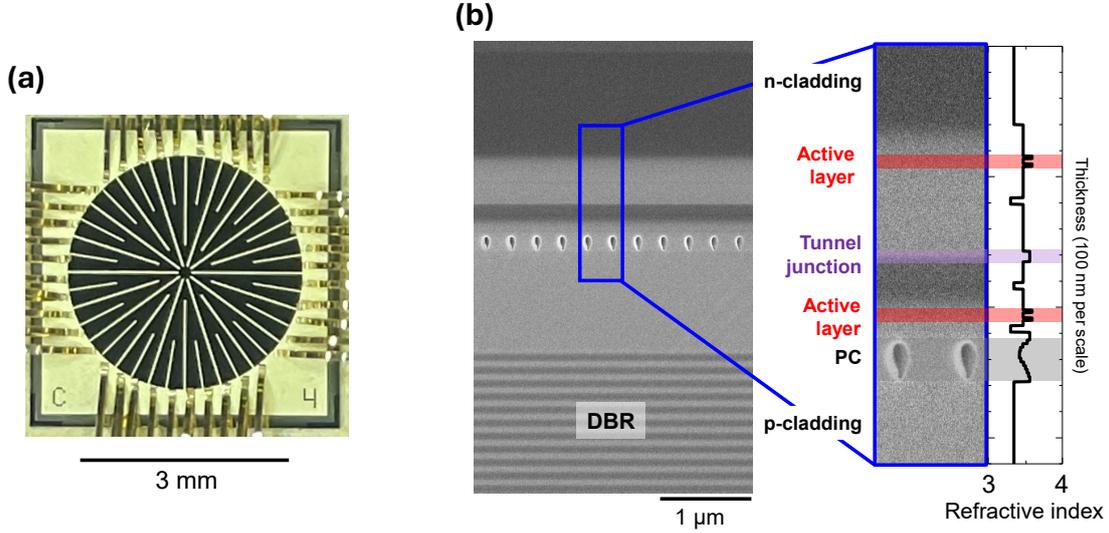

**Fig. 3.** Fabricated two-junction PCSEL. (a) Microscope image of the 3-mm-diameter two-junction PCSEL. (b) Cross-sectional SEM image of the two-junction PCSEL.

Figure 4(a) shows the photonic band diagram measured below the threshold current and the lasing spectrum measured above the threshold current of the fabricated two-junction PCSEL. Band edges originating from the first higher-order waveguide mode are clearly visible as indicated in Fig. 4(a), owing to the low $\Gamma_{tunnel}$ of this mode, as well as to the matching of the $\Gamma$-point band-edge frequencies of this mode to the gain peak wavelength as described above via adjustment of the photonic crystal lattice constant. On the other hand, the luminescence of the band edges corresponding to the fundamental waveguide mode (whose presence were detected by enhancing the contrast of the image as shown in the inset) were weak owing to the mismatch of band-edge frequencies and the gain peak wavelength, as well as by the large absorption loss caused by the tunnel junction. These measurements confirmed that lasing oscillation occurred at the band edge (specifically, band-edge A) of the first higher-order waveguide mode as designed. From the measured photonic band structure, the coupling coefficients were estimated to be $R$=49 cm$^{-1}$, $I$=30 cm$^{-1}$ and $\mu$=50 cm$^{-1}$. The radiation coefficient $\alpha_v$ derived from this estimation is sufficiently large at 8.5 cm$^{-1}$, and thus operation with a high slope efficiency is expected. It should be noted that the estimated $R$, $I$ and $\mu$ are slightly different from the designed values, and thus there remains room to optimize $R$, $I$ and $\mu$ to achieve higher beam qualities.

Next, we mounted the two-junction PCSEL onto a driver circuit for high-current pulse driving and measured the *I-L* characteristics and far-field patterns (FFPs). We set the pulse width and repetition frequency to 25 ns and 200 Hz, respectively. Figure 4(b) shows the measured *I-L* characteristics of the two-junction PCSEL together with those of a conventional 3-mm single-junction PCSEL for reference. The two-junction PCSEL had a slope efficiency of 1.58 W/A, which is over twice as high as that of the single-junction PCSEL, and a peak output power of 1.8 kW, limited by the maximum diver circuit current of 1200 A. This peak output power is approximately twice that of a 3-mm-diameter single-junction PCSEL, and the highest value ever achieved with a PCSEL. As shown in Fig. 4(c), a narrow, single-lobed beam of 0.32° in the x direction and 0.31° in the y direction ($1/e^2$ width) was obtained under a high injection current of 1200 A (65 times the threshold current). The laser brightness, evaluated using the measured peak output power and FFP widths at 1200 A, was 1 GW cm$^{-2}$ sr$^{-1}$. As mentioned above, the beam quality can be improved by optimizing the balance and magnitude of the coupling coefficients *R*, *I*, and *μ*, whereupon an even higher brightness can be realized.

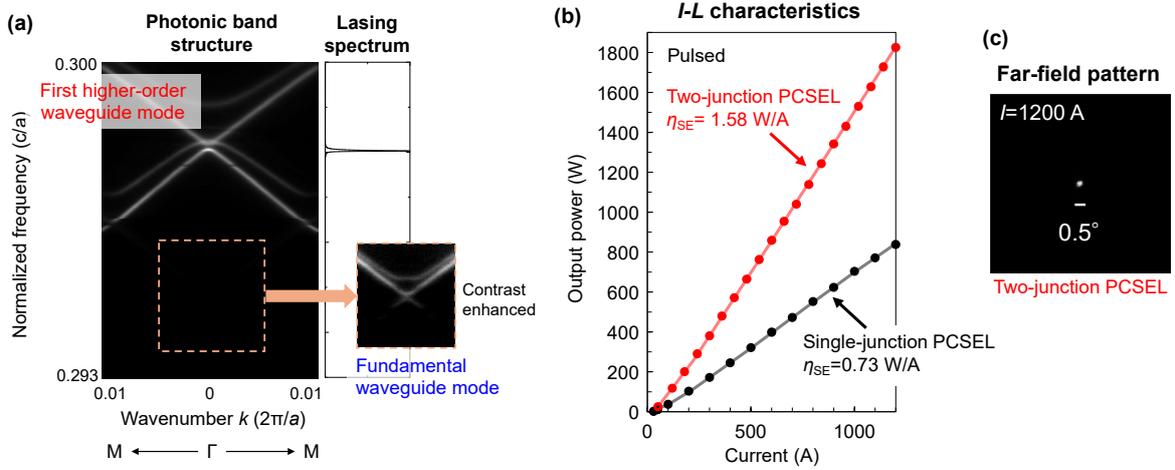

**Fig. 4.** Measured characteristics of the fabricated 3-mm-diameter two-junction PCSEL under pulsed operation. (a) Photonic band diagram and lasing spectrum. (b) *I-L* characteristics of a 3-mm two-junction PCSEL (red). *I-L* characteristics of a 3-mm-diameter single-junction PCSEL are also shown for reference (black). (c) Far-field pattern at an injection current of 1200 A.

In summary, we have investigated PCSELs with multijunction active layers for the purpose of increasing the slope efficiency, and we have designed a layer structure of a two-junction PCSEL, which operates in the first higher-order waveguide mode to minimize internal material losses and obtain sufficient optical gain and resonance effects. In addition, we have simulated the lasing characteristics of the two-junction PCSEL, showing that a slope efficiency almost twice as high as that of conventional single-junction PCSEL can be expected while maintaining a high-beam quality. Next, we have fabricated a 3-mm-diameter two-junction PCSEL based on the above design and have evaluated the lasing characteristics of this PCSEL under pulsed operation, achieving a slope efficiency of 1.58 W/A and a peak output power of 1.8 kW while maintaining a high brightness of 1 $GWcm^{-2}sr^{-1}$. This slope efficiency is over twice as high as that of conventional single-junction PCSELs, and this peak output power is the highest ever achieved with a PCSEL. By increasing the number of active and tunnel-junction layer pairs, an even higher slope efficiency is expected. Our proposed multijunction active layer can also be applied to M-PCSELs, which operate in a band edge at the M point of the photonic crystal and enable beam scanning and the emission of various beam patterns, to improve the maximum detection range of these light sources in LiDAR applications.

**Acknowledgments**

This work was partially supported by the project of the Council for Science, Technology and Innovation (Program for Bridging the Gap between R&D and the Ideal Society (Society 5.0) and Gathering Economic and Social Value (BRIDGE)). This work was also partially supported by Grant-in-Aid for Scientific Research (22H04915) of the Japan Society for the Promotion of Science, and by the project (JPNP22007) of the New Energy and Industrial Technology Development Organization (NEDO).